\documentclass[11pt]{article}

\usepackage[utf8]{inputenc}
\usepackage[T1]{fontenc}
\usepackage{amsmath}
\usepackage{parskip}
\usepackage[margin=1in]{geometry}
\usepackage{graphicx}
\usepackage{float}
\usepackage{booktabs}
\usepackage{hyperref}
\usepackage[round]{natbib}

\title{Towards an LLM-based method for quantifying the sexual content in song lyrics}
\author{Ignacio M. Sticco}
\date{\today}

\begin{document}
\maketitle

\begin{abstract}
Reggaeton is one of the most widely consumed music genres in the world, and its
lyrics are commonly regarded as highly sexualized. This claim rests mostly on
qualitative studies and on small-scale quantitative ones. This paper has two
goals. First, we present a reproducible method that uses a large language model
to quantify thematic content in song lyrics along several independent
dimensions. The method is not restricted to sexual content. Second, we apply it
to a corpus of 1{,}259 songs by 12 reggaeton artists released between 2002 and
2025. The analysis covers four topics: a dataset characterization, a per-artist
comparison, an analysis of how the dimensions change over time, and a
comparison between our
sexual-explicitness score and Spotify's own \emph{explicit} flag. We release the
data collection code, the scoring prompt, and the corpus, so that other
researchers can replicate the approach or apply it to their own lyrics
datasets.
\end{abstract}

\section{Introduction}
\label{sec:introduction}

Reggaeton is one of the most widely consumed music genres in the world, and the
most listened-to genre in Spanish \citep{merlyn2020dime}.
Few things about it are repeated as often as the observation that its lyrics
are saturated with sexual content. This is not only a popular impression.

\citet{carballo2006musica} described the genre early on as a social phenomenon
whose lyrics combine sexual content with symbolic violence.
\citet{hellin2021sex} analyzed 118 songs by 25 male performers and found sex to
be the main thread of reggaeton discourse. The mechanism is a metaphorical
animalization of both genders, as predators and prey. The two are not
symmetric, however. The animalization of women operates at the service of men,
and the author ties it to sexual objectification and dehumanization. A related
line of work has examined the genre in terms of gender violence
\citep{arevalo2018diabla}.

The genre is by now a well-established object of study, and two recent
bibliometric reviews give a good picture of the field.
\citet{arango2025reggaeton} surveyed 72 Scopus-indexed articles published
between 2006 and 2024, and found that gender stereotypes account for 38\% of
that literature. Song lyrics are among the most frequently studied units of
analysis, and only 6 of the surveyed articles follow a purely quantitative
design. Lyrics are not the only object of study, however: music videos are
analyzed as well, and \citet{merlyn2020dime} coded the videos and the lyrics of
the 100 most popular reggaeton songs of 2018 side by side.
\citet{espinal2024tendencias} reach a similar diagnosis across 69 articles. The
former review observes that the techniques and instruments used to study
reggaeton remain largely unconsolidated, with few studies reusing the procedures
of earlier ones.

That observation is the starting point of this paper. The interpretive tradition
has done the conceptual work of identifying which themes are at play in the
genre and what they mean. Once those themes are named, a second family of
questions follows: how much of each theme is actually present, how it has
changed over time, and how much artists within the same genre differ from
one another.

These are questions about magnitude, and answering them requires measurement. By
measurement we mean an explicit, documented procedure that assigns comparable
numbers to songs. Another researcher must be able to run that procedure again,
on the same corpus or on a different one. We see such a measurement layer as
complementary to the interpretive work that defined the constructs in the first
place. Quantification turns claims about a genre into claims that can be
compared across artists, tracked over time, and revised in light of new data.

What has limited measurement at scale is the cost of coding songs by hand.
\citet{arevalo2018diabla} coded 70 songs, awarding one point per mention of each
of five types of gender violence. \citet{rodriguez2024violencia} coded 20 of the
most-streamed songs in Mexico using binary indicators, over categories adapted
from \citet{bretthauer2007feminist}. That earlier feminist content analysis of
popular music distilled six recurring themes: men and power, sex as a male
priority, objectification of women, sexual violence, women defined by having a
man, and women not valuing themselves. \citet{diez2022inequality} applied
software-assisted thematic analysis to the 65 most commercially successful
reggaeton songs of 2020, identifying which themes are present rather than how
much of each one is. Related work draws on the distinction between hostile and
benevolent sexism introduced by \citet{glick1996ambivalent}.

The constructs these studies use are well established. What constrains them is
that every additional song costs human labor, so the number of songs analyzed
and the depth of the coding must be traded off against one another.

Computational approaches relax that constraint in different ways.
\citet{torres2023computational} retrieved 641 songs by twelve female artists and
analyzed them through tokenization and word-frequency techniques. That approach
scales easily, but it operates on individual words rather than on themes.

Supervised classifiers work at the level of meaning.
\citet{damian2025misongyny} treat each song as a bag of sentences and apply
multiple instance learning to a shared-task corpus of 2{,}104 Spanish-language
songs for misogyny detection. \citet{zamacola2026finetuning} fine-tune a GPT
model on 100 expert-labeled songs to classify Spanish-language lyrics as
sexually explicit or not, with a view to automated moderation and age-based
rating on streaming platforms. Both approaches require labeled training data.
Moreover, both return categorical labels rather than graded scores across
several themes at once.

A third possibility has emerged more recently: using a large language model
directly as a rater.\footnote{Modern LLMs are built on the transformer
architecture introduced by \citet{vaswani2017attention}.}
\citet{zhang2024positive} frame music content assessment as
a multi-task problem over five aspects (violence, substance use, sex,
consumerism, and positive messages), rated on ordinal severity levels. They also
include case studies probing where language models over- and underestimate each
aspect. \citet{dahary2025joy} build a human-annotated benchmark of emotion
intensity in pop lyrics and evaluate several LLMs zero-shot against it.
\citet{chandra2025longitudinal} apply language models to Billboard chart lyrics
across seven decades, and report a marked rise in explicit content from 1990
onward.

Taken together, this line of work suggests that language models can produce
graded, multi-dimen\-sional ratings of lyrics at scale, and that such ratings
support longitudinal analysis. To our knowledge, this has not yet been brought
to bear on reggaeton with a scoring protocol that is calibrated against human
labels and released for reuse.

This paper makes two contributions. The first one is methodological: we present
a reproducible procedure that takes a list of artists and returns a table in
which each row is a song and each column a thematic dimension scored by an LLM.
Nine dimensions are scored on an ordinal scale, among them a deliberate
separation between sexually \emph{suggestive} and sexually \emph{explicit}
content, and the scoring prompt is calibrated against hand-labeled songs before
being applied to the corpus.

The second contribution is empirical. We apply the method to 1{,}259 reggaeton
songs by 12 artists released between 2002 and 2025. To the best of our
knowledge, this is the largest reggaeton corpus analyzed at the level of lyrical
content, next to corpora ranging from 20 to 641 songs in the studies cited
above. We report a dataset characterization, a per-artist comparison, a
longitudinal analysis of how the dimensions change over the two decades
covered, and a comparison between
our sexual-explicitness score and Spotify's own \emph{explicit} flag.

Nothing in the method is specific to sexual content. The dimensions are defined
in the scoring prompt. Therefore, the same procedure could be pointed at the six
themes of \citet{bretthauer2007feminist}, at the hostile and benevolent sexism
of \citet{glick1996ambivalent}, or at any other coding scheme a researcher wants
to apply at scale. It could also be pointed at a different genre or a different
language. We release the data collection code, the scoring prompt, and the
corpus itself so that the procedure can be replicated, audited, and adapted. We
hope it makes quantitative questions about song lyrics cheaper to ask.

The paper is organized as follows. In the following section, we describe the
three steps of the method and its limitations. In Section~\ref{sec:results}, we
present the results of applying it to the reggaeton corpus. In the last part, we
summarize the main conclusions of the work.

\section{Method}
\label{sec:method}

The goal of the method is to go from a list of artists to a structured corpus.
Each row of that corpus is a song, and each column is a thematic dimension
scored by a large language model (LLM). The full implementation is released as
an open-source repository, so that other researchers can reproduce it or adapt
it to a different corpus.\footnote{% repo will be made public once this pre-print is released
\url{https://github.com/ignaciosticco/open-lyrics-scorer}}

The method consists of three steps. In the first step (\emph{data collection}),
lyrics and track metadata are gathered and manually reviewed. In the second step
(\emph{LLM-based scoring}), each song is scored along several thematic
dimensions. In the third step (\emph{corpus assembly}), per-song scores are
normalized and deduplicated into a single analysis-ready table.

\subsection{Step 1: Data collection}

For each artist, we collect every track from their studio albums via the
Spotify Web API (Client Credentials flow). For each track we retrieve the
metadata (album, release year, duration, and Spotify's own \emph{explicit}
flag), together with the corresponding lyrics from the Genius API. Only full
albums are considered, with no singles or compilation appearances. Therefore,
every collected track is unambiguously attributed to a single artist: the artist
credited on the album, and not necessarily every featured artist on the track.

Genius does not have lyrics for every song. Some entries exist but have no
lyrics transcribed, and these cases are marked automatically. Separately, some
of the lyrics that Genius does return are contaminated with content that is
clearly not part of the song (page metadata, annotations, or other boilerplate
text). That content is removed via a combination of regular-expression rules and
manual inspection.

Both issues are addressed through a deliberate manual review step rather than a fully
automated one. Recovering a missing or corrupted lyric requires judgment calls,
such as matching the correct version of a song, skipping instrumental tracks and
skits, or telling genuine lyrics apart from spurious text. When a song is
marked, we cross-check it against alternative lyrics sites before including or
excluding it from the corpus. The sites we used are \url{https://www.letras.com},
\url{https://www.musica.com}, and \url{https://www.musixmatch.com}.

This makes the data collection step semi-automatic rather than fully automatic.
The cost of the manual review is bounded, however, because only a minority of
tracks come back from Genius missing or contaminated. The automatic retrieval
resolves most of the corpus on its own, and manual attention is spent only on
the flagged remainder. Assembling a corpus of this size song by song would take
substantially longer.

\subsection{Step 2: LLM-based scoring}

\begin{figure}[H]
\centering
\includegraphics[width=0.85\textwidth]{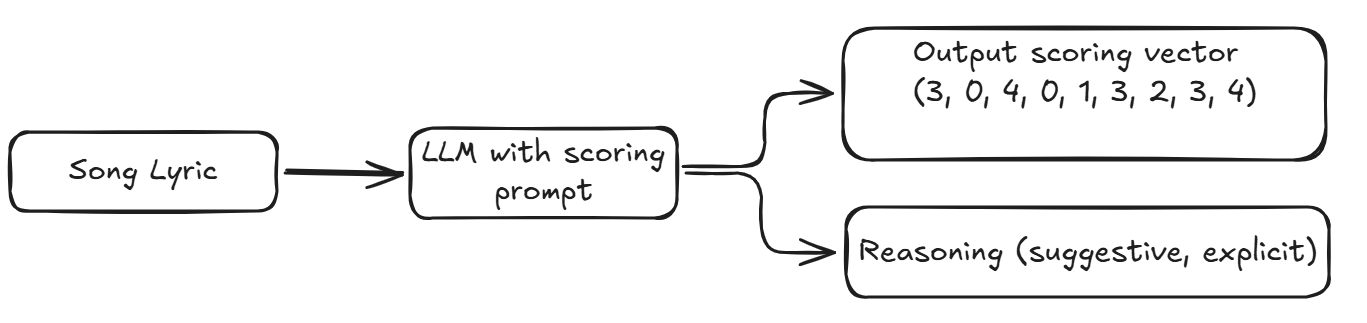}
\caption{Scoring pipeline for Step 2. Each song's lyrics are passed
individually to an LLM together with the scoring prompt. The model returns a
nine-element score vector (one score per thematic dimension), plus a written
reasoning field for each of the two sexual dimensions.}
\label{fig:scoring-pipeline}
\end{figure}

Each song's lyrics are scored individually by \texttt{gpt-5.1}
\citep{openai2025gpt51}, one song per call. The corpus was scored between May
and June 2026. Nine independent thematic dimensions are scored:
\emph{party/nightlife},
\emph{substance use}, \emph{wealth/status}, \emph{street crime},
\emph{infidelity}, \emph{romantic emotion}, \emph{personal reflection},
\emph{sexual suggestiveness}, and \emph{sexual explicitness}. Each dimension
is scored on a 5-point ordinal scale (0 = absent to 4 = dominant theme). The
score is based on thematic frequency, centrality, and prominence within the song
as a whole, rather than on isolated lines. Together, these nine per-dimension
scores form the song's output scoring vector, as illustrated in
Figure~\ref{fig:scoring-pipeline}.

A key methodological distinction is drawn between \emph{suggestive} and
\emph{explicit} sexual content. Suggestive content evokes sexuality through
metaphor or connotation, without naming it directly. Explicit content, instead,
names sexual acts, body parts, or behavior directly, regardless of register
(vulgar or otherwise). The scoring
prompt provides anchor examples for each point on the scale, in order to keep
this distinction consistent across songs.

Arriving at a prompt that reliably matches human judgment is an iterative
process, and we treat it as a central part of the method rather than an
incidental one. This is particularly relevant for researchers who wish to apply
the method with a different scoring criterion, or to a different genre.

In practice, the calibration works as follows. We draw a small calibration
sample of about ten songs at random, and label them by hand along all nine
dimensions. We then compare those labels against the LLM's output under the
current prompt. Discrepancies are used to revise the prompt, by clarifying a
definition or adding an anchor example. The cycle is repeated until the LLM's
scores are consistently aligned with the manual labels.

A held-out test set of five additional songs, not used during calibration, is
then scored to confirm that the alignment generalizes beyond the calibration
sample. We recommend this calibration procedure to anyone adapting the method
to a new criterion, language, or genre, since a prompt tuned for one context
need not transfer to another.

For the two sexual dimensions, the prompt additionally requests a short written
justification alongside the numeric score. These justifications are stored as
two auxiliary fields in the output. They are not used in the quantitative
analysis, but they are essential for auditing the scores. This is particularly
useful during prompt calibration, when disagreements between the LLM and the
manual labels are easier to diagnose by reading the model's stated reasoning
than by inspecting the score alone.

\subsection{Step 3: Corpus assembly}

Scores are rescaled from the original 0--4 ordinal range to a 0--1 range for
analysis. Throughout the rest of the paper, \emph{score} always refers to this
normalized 0--1 value.

Two duplicate-detection passes are applied. The first pass handles
within-artist duplicates (say, a live or remix version of a song already
present in its studio form). These are resolved by preferring the studio
version and, if tied, the earliest release year. The second pass handles
cross-artist duplicates, that is, the same song legitimately collected once per
collaborating artist on a joint album or a feature. These are kept as separate
rows, one per credited artist, but their scores are averaged across the group.
Therefore, the same song does not receive two different ratings depending on
whose page it was collected from.

A \emph{composite sexual score} is derived as the average of the two sexual
dimensions:
\[
  \text{sexual}_{\text{composite}} = \frac{\text{explicit}_{\text{norm}} + \text{suggestive}_{\text{norm}}}{2}.
\]

Full implementation details for all three steps are available in the
\texttt{open-lyrics-scorer} repository referenced above. These include the exact
scoring prompt, the deduplication logic, and the data collection scripts.

\subsection{Limitations and future work}

We note three limitations of the method. In the first place, different LLMs can
produce noticeably different scores for identical lyrics. The same happens
across different versions of the same model family. Scores should therefore be
treated as specific to the model version used, and not blindly compared across
studies that use a different one. Note also that \texttt{gpt-5.1} is a model
alias rather than a pinned snapshot. The provider may repoint it over time, so
an exact numerical replication is not guaranteed even with the same prompt.

Secondly, the method inherits any biases the underlying LLM has around
language, dialect, or cultural context. This is of particular relevance for
this corpus, given its heavy use of Caribbean Spanish slang and code-switching.

Finally, the method was developed and validated on Spanish-language lyrics. We
expect the underlying approach to extend to other languages. Nevertheless, this
has not been tested directly, and it would require its own calibration cycle.

Looking forward, two directions could make the method both more accurate and
more automatic. The first one is to replace the general-purpose LLM with a model
fine-tuned specifically for this kind of scoring task, which could improve
accuracy and consistency. This follows the approach of
\citet{zamacola2026finetuning} for explicit-content detection in
Spanish-language lyrics. The second one is to automate part of the manual review
step of Step 1. An LLM could first judge whether a retrieved lyric is complete
and free of spurious content. If it is not, the LLM could automatically query
alternative lyrics providers, rather than relying on manual cross-checking. We
leave both directions to future work, by us or by other researchers building on
the released method.

\section{Results}
\label{sec:results}

This section applies the method to reggaeton and reports four analyses of the
resulting corpus: a characterization of the corpus, a comparison between
artists, a longitudinal analysis, and a comparison against Spotify's explicit
flag.

\subsection{Dataset characterization}
\label{sec:dataset-characterization}

Before comparing artists or years, we describe the corpus itself. This
subsection covers its size and time span, the distribution of each of the nine
dimensions, and the correlation structure between them.

\begin{figure}[H]
\centering
\includegraphics[width=0.9\textwidth]{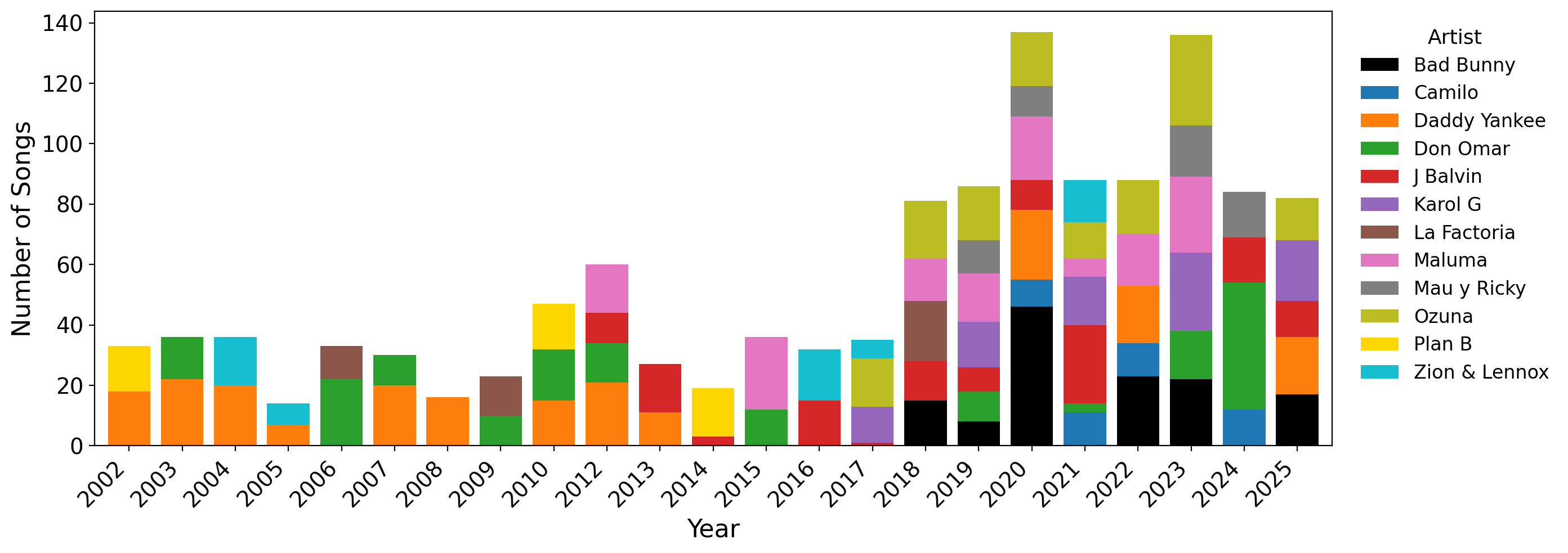}
\caption{Number of songs per year, colored by artist. Bars are stacked, so the
height of each bar is the total number of songs collected for that year.}
\label{fig:corpus-timeline}
\end{figure}

Applying the method to our list of 12 reggaeton artists yields a corpus of
1{,}259 songs spanning 2002--2025, with a single missing year (2011), in which
none of the sampled artists released studio material.
Figure~\ref{fig:corpus-timeline} shows the number of songs per year and artist.
The corpus is heavily weighted toward recent years, with 62\% of songs released
from 2018 onward. This reflects both the artists' actual output and the growth
of streaming-era catalogs. Notice that recent years are populated by a partly
different subset of artists than early years. This uneven coverage is worth
keeping in mind when interpreting the longitudinal analysis below.

\begin{figure}[H]
\centering
\includegraphics[width=0.55\textwidth]{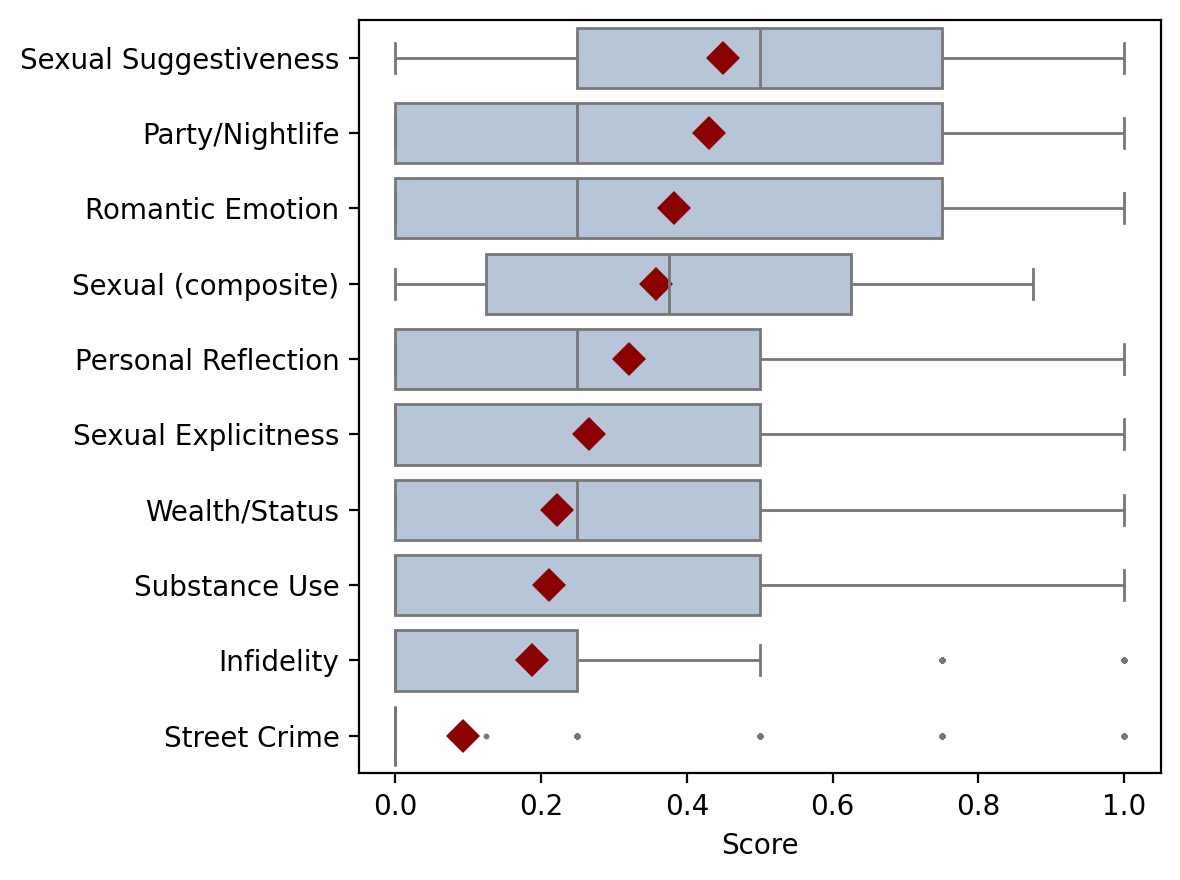}
\caption{Distribution of all nine thematic dimensions and the composite
sexual score, ordered by mean. Boxes span the interquartile range, the inner
line is the median, and the whiskers extend to 1.5 times that range. Points
beyond the whiskers are individual songs falling outside it. Diamonds mark the
mean.}
\label{fig:dims-boxplot}
\end{figure}

Figure~\ref{fig:dims-boxplot} summarizes the distribution of all nine thematic
dimensions plus the composite sexual score. \emph{Sexual suggestiveness} has the
highest mean of the nine individual dimensions (0.45), followed by
\emph{party/nightlife} and \emph{romantic emotion}. \emph{Street crime} and
\emph{infidelity} are the least present. Notice that \emph{sexual
suggestiveness} is nearly twice as prevalent as \emph{sexual explicitness}
(0.27). This is an early sign that the suggestive/explicit distinction
introduced in Step 2 captures a real asymmetry in the corpus, rather than a
redundant pair of labels.

\begin{figure}[H]
\centering
\includegraphics[width=0.75\textwidth]{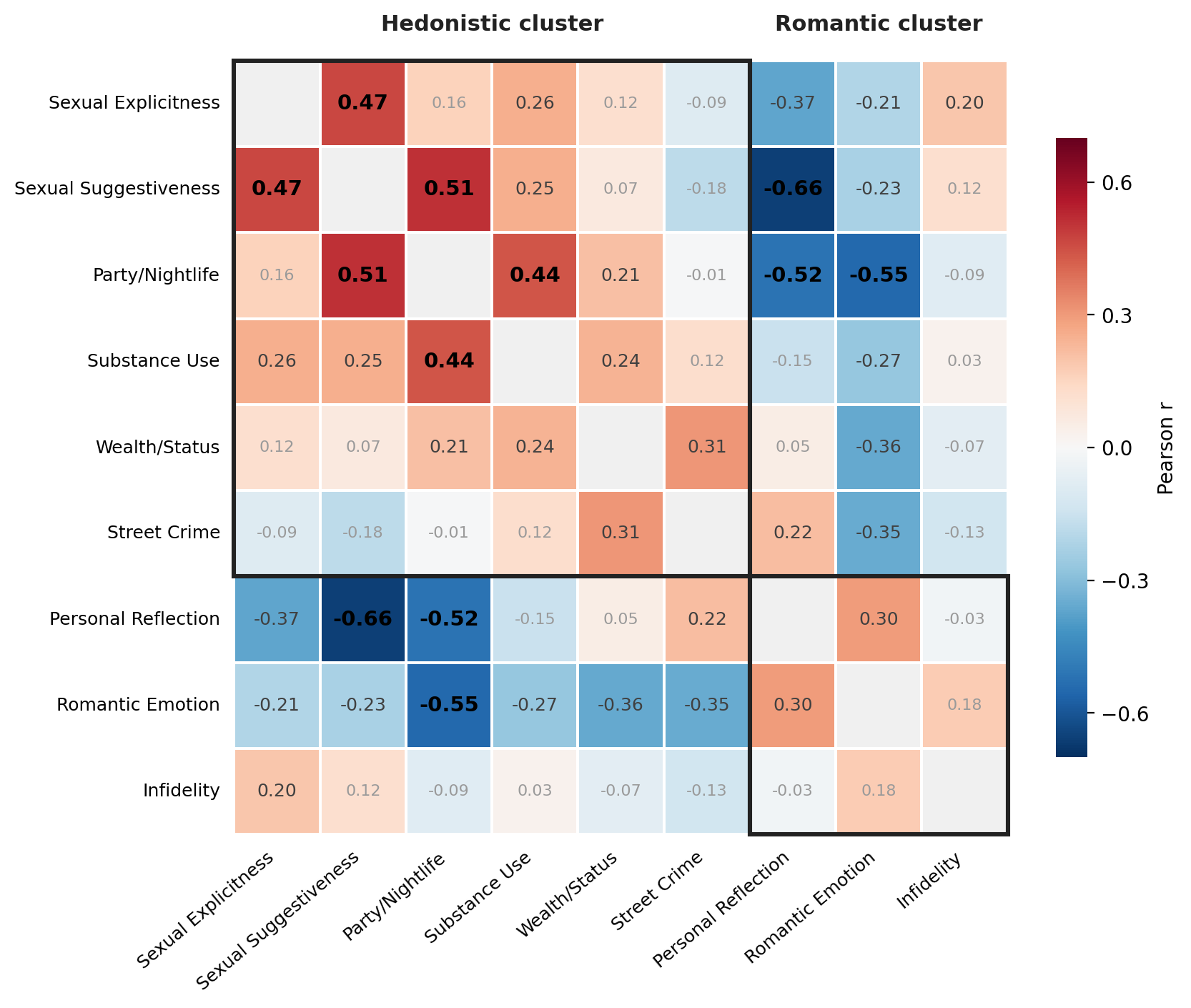}
\caption{Pairwise correlations between all nine thematic dimensions. Rows and
columns are ordered by hierarchical clustering. The boxes mark the two clusters
obtained by cutting the dendrogram in two, labeled \emph{Hedonistic cluster}
and \emph{Romantic cluster}. The composite sexual score is omitted here because it is an
average of two of the dimensions shown, so its correlations with them follow by
construction.}
\label{fig:corr-matrix}
\end{figure}

Figure~\ref{fig:corr-matrix} shows pairwise correlations across dimensions.
Throughout the paper, $r$ denotes the Pearson correlation coefficient.
Rows and columns are ordered by hierarchical clustering (average linkage on
$1-r$), so that co-varying dimensions sit adjacent. Two loose thematic poles
emerge.

On one side, \emph{party/nightlife}, \emph{substance use}, and \emph{sexual
suggestiveness} are all positively correlated with one another
(\emph{Hedonistic cluster} in Figure~\ref{fig:corr-matrix}). Party/nightlife
correlates $r=0.44$ with substance use, and $r=0.51$ with sexual
suggestiveness.
On the other side, \emph{personal reflection} and \emph{romantic emotion} are
the two dimensions most consistently anti-correlated with that cluster
(\emph{Romantic cluster}).
Personal reflection has the strongest correlation in absolute value of the
whole matrix, with \emph{sexual suggestiveness} ($r=-0.66$), and it also has a
sizeable negative correlation with \emph{sexual explicitness} ($r=-0.37$).
Romantic emotion, in turn, is strongly anti-correlated with party/nightlife
($r=-0.55$).

\emph{Sexual suggestiveness} and \emph{sexual explicitness} are themselves
positively correlated ($r=0.47$), but well short of $r=1$. Therefore, they track
overlapping sets of songs without being interchangeable measurements. This is
what makes the distinction introduced in Step 2 worth maintaining.

\subsection{Per-artist analysis}
\label{sec:per-artist}

This subsection examines how much artists differ from one another. We compare
them first on the composite sexual score, and then on their full profile across
dimensions.

\begin{figure}[H]
\centering
\includegraphics[width=0.9\textwidth]{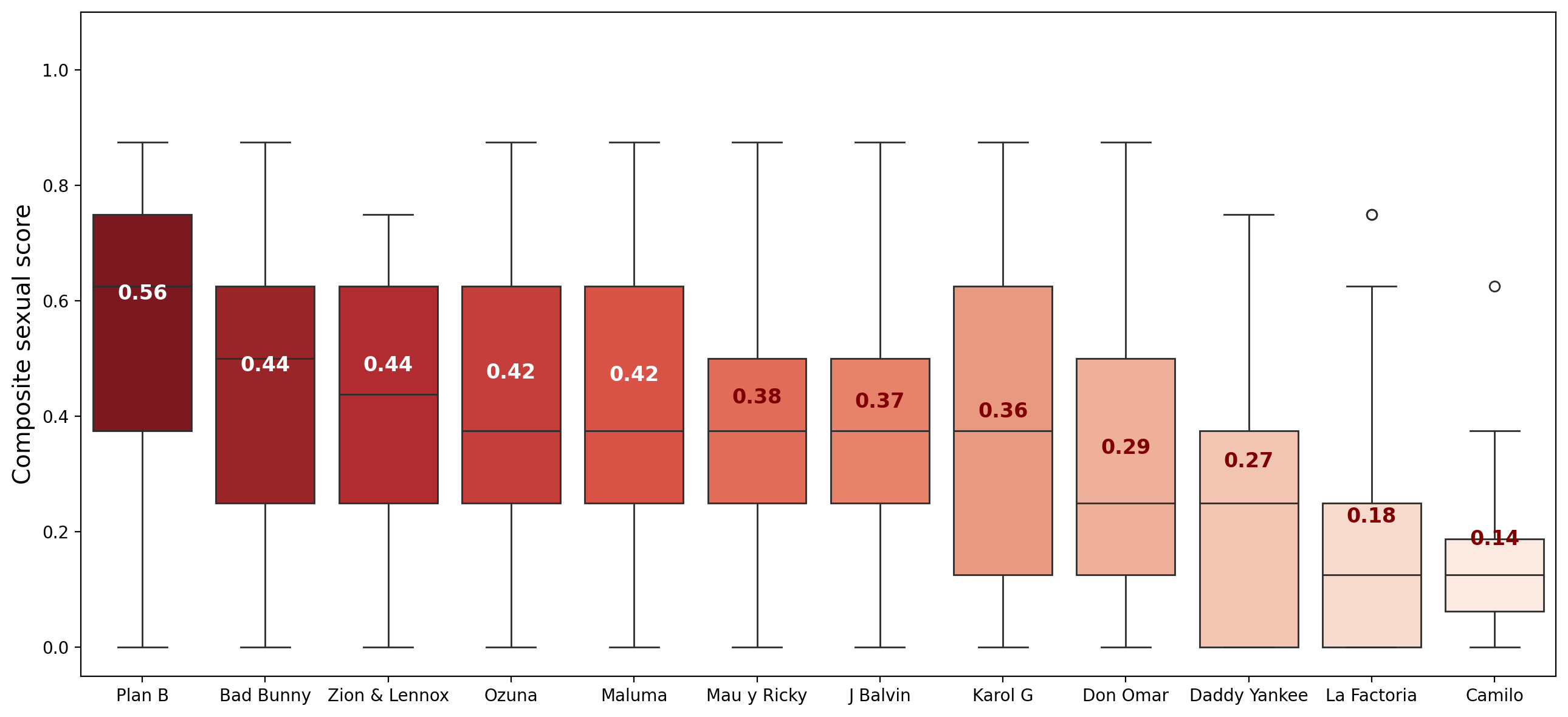}
\caption{Distribution of the composite sexual score by artist, ranked by mean
(labeled). The 12 artists of the corpus are shown. Boxes, whiskers, and outlier
points follow the same convention as Figure~\ref{fig:dims-boxplot}.}
\label{fig:artist-boxplot}
\end{figure}

Figure~\ref{fig:artist-boxplot} ranks the 12 artists by mean composite sexual
score. The range is wide: Plan B tops the ranking (mean 0.56), while Camilo
sits at the opposite end (mean 0.14). This is a four-fold difference between
the two extremes. Daddy Yankee ranks comparatively low (0.27), despite being
one of reggaeton's founding and most commercially prominent figures. Bear in
mind that sample sizes vary considerably across artists, from 43 to 211 songs.
The per-artist counts are not shown in the figure, and can be consulted in the
released corpus.

\begin{figure}[H]
\centering
\includegraphics[width=\textwidth]{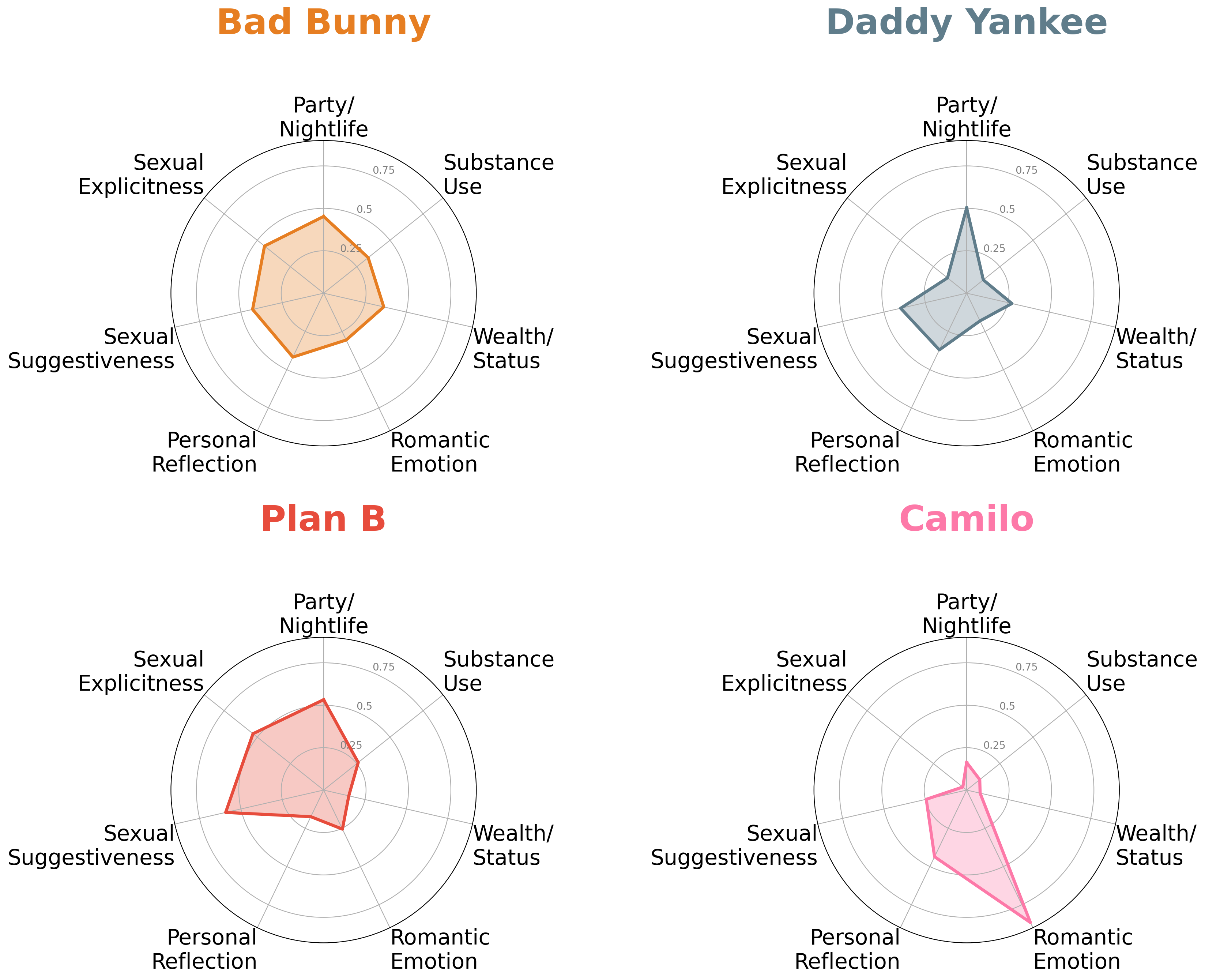}
\caption{Mean score on seven thematic dimensions for four representative
artists. Each vertex is the mean over all of that artist's songs, and the four
panels share the same radial scale, from 0 to 0.9.}
\label{fig:radar}
\end{figure}

To look beyond the single composite sexual score, Figure~\ref{fig:radar} plots
four artists across seven of the nine thematic dimensions. The four artists were
chosen to span the range observed above. \emph{Street crime} and
\emph{infidelity} are omitted for readability, since both are rare overall. The
four profiles read as distinct archetypes, rather than variations on a theme.

Bad Bunny is the most evenly spread of the four, with all seven dimensions
falling in a comparatively narrow 0.31--0.45 range. This is consistent with a
catalog that draws on many themes rather than specializing in one. Daddy Yankee
is dominated by party/nightlife (0.50), with sexual suggestiveness the
next-highest dimension and sexual explicitness notably low (0.15), which reads
as a ``party anthem'' profile more than a sexually explicit one. Plan B is
dominated by the two sexual dimensions, sexual suggestiveness (0.59) and sexual
explicitness (0.53), which are the two highest single values among the four
artists shown. Plan B also has comparatively little wealth/status or personal
reflection content. Camilo is the most extreme case in the whole corpus:
romantic emotion (0.87) dwarfs every other dimension, while sexual explicitness
is nearly absent (0.03).

\begin{figure}[H]
\centering
\includegraphics[width=0.85\textwidth]{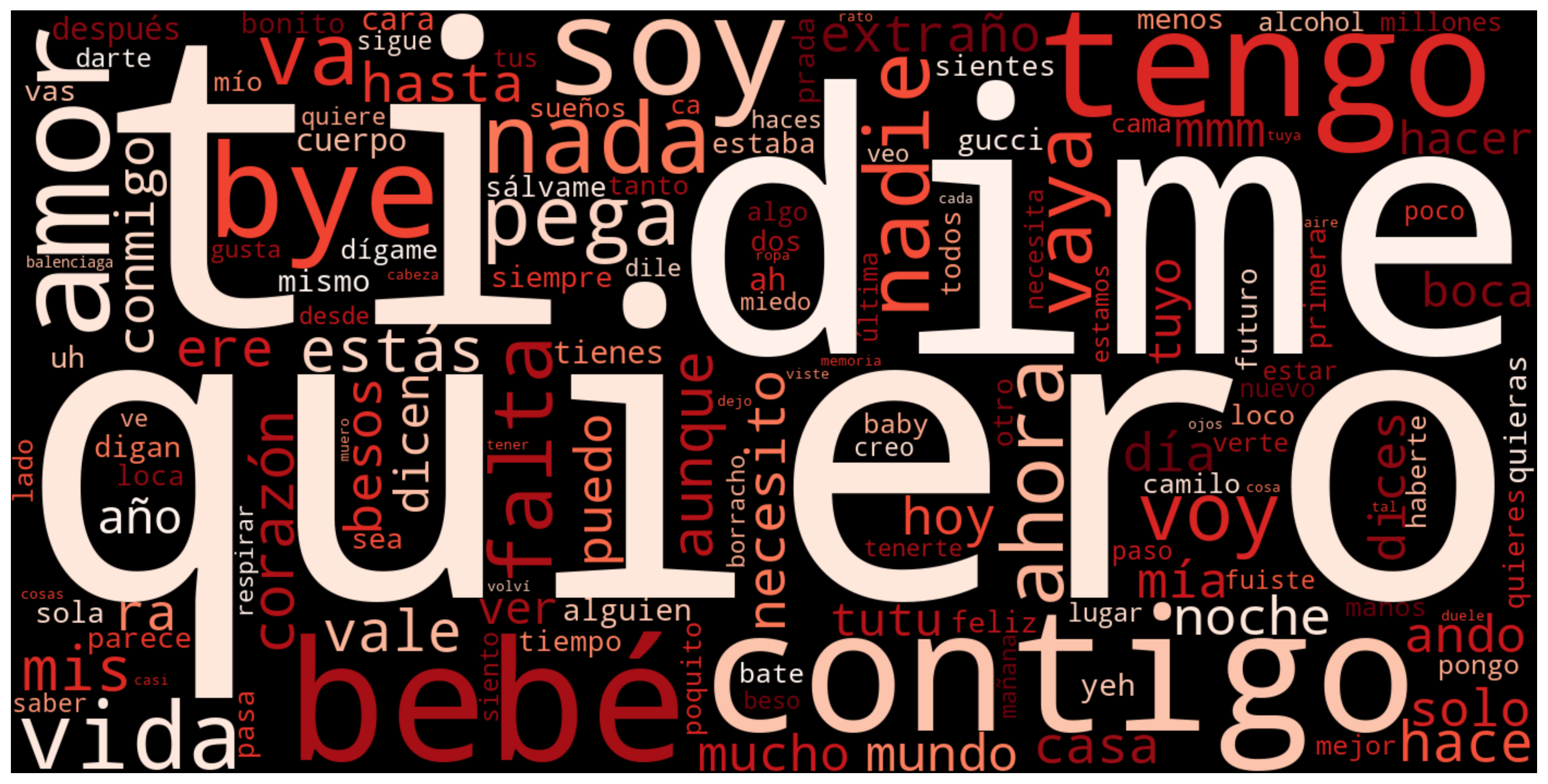}
\caption{Word cloud from the lyrics of Camilo (n=43 songs). Word size is
proportional to raw frequency, and the 150 most frequent words are shown.
Spanish and English stopwords are removed, such as \emph{que}, \emph{de},
\emph{en}, and \emph{la}.}
\label{fig:wc-camilo}
\end{figure}

\begin{figure}[H]
\centering
\includegraphics[width=0.85\textwidth]{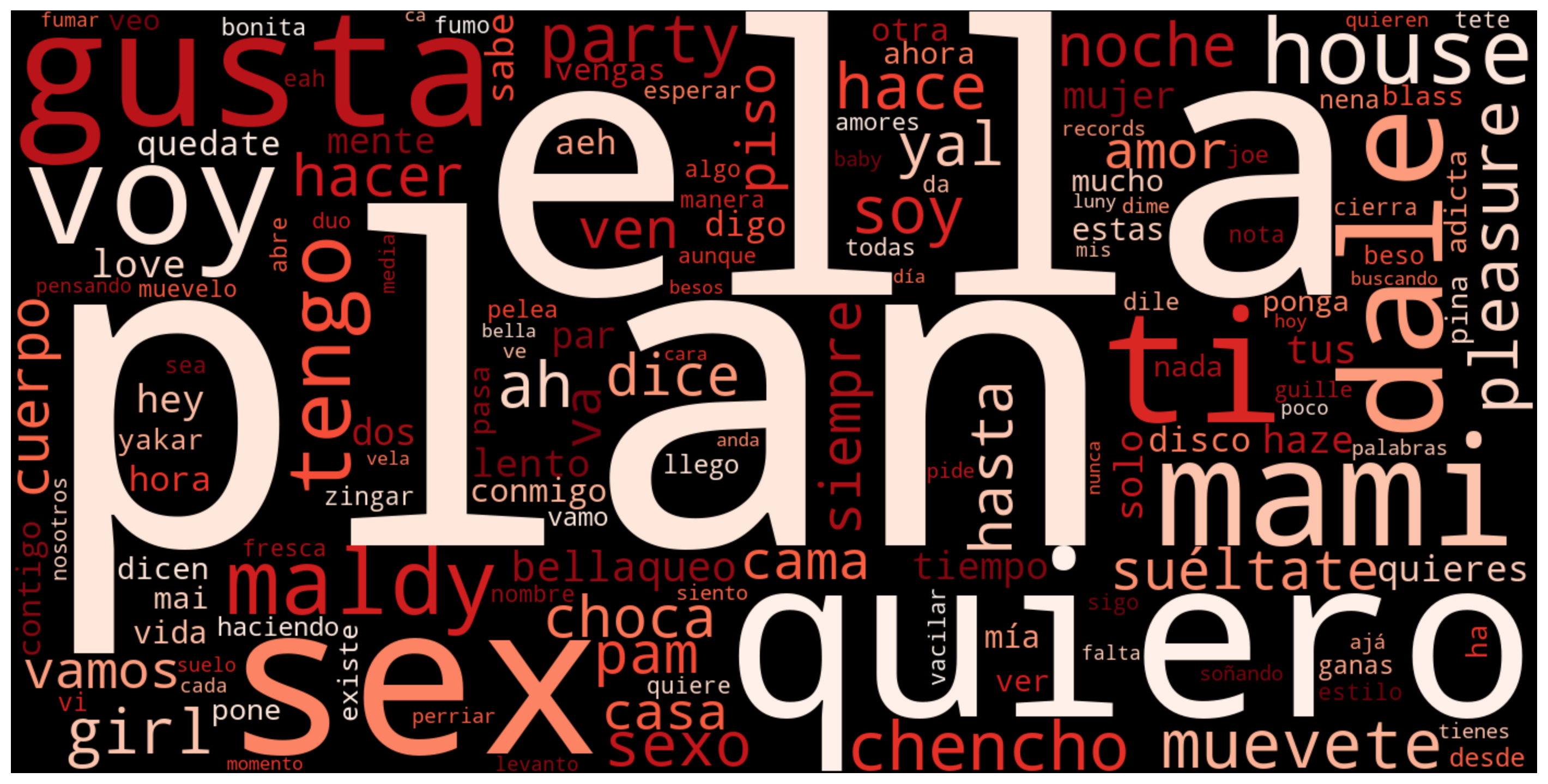}
\caption{Word cloud from the lyrics of Plan B (n=46 songs). Built in the same
way as Figure~\ref{fig:wc-camilo}.}
\label{fig:wc-planb}
\end{figure}

Camilo and Plan B sit at opposite ends of the composite sexual score ranking.
Word clouds built from their raw lyrics offer a qualitative check on these
quantitative profiles (Figures~\ref{fig:wc-camilo} and~\ref{fig:wc-planb}).
Camilo's most frequent words are overwhelmingly romantic and tender
(\textit{amor}, \textit{quiero}, \textit{dime}, \textit{contigo},
\textit{bebé}, \textit{vida}). Plan B's include direct sexual vocabulary
(\textit{sex}, \textit{sexo}, \textit{cuerpo}, \textit{mami}, \textit{cama},
\textit{bellaqueo}). The contrast is visible at a glance, without appealing to
any score, and it lines up with the quantitative picture above. The LLM-based
scores are not just numbers. They track a real difference in the words artists
actually use.

\subsection{Longitudinal analysis}
\label{sec:longitudinal}

\begin{figure}[H]
\centering
\includegraphics[width=0.85\textwidth]{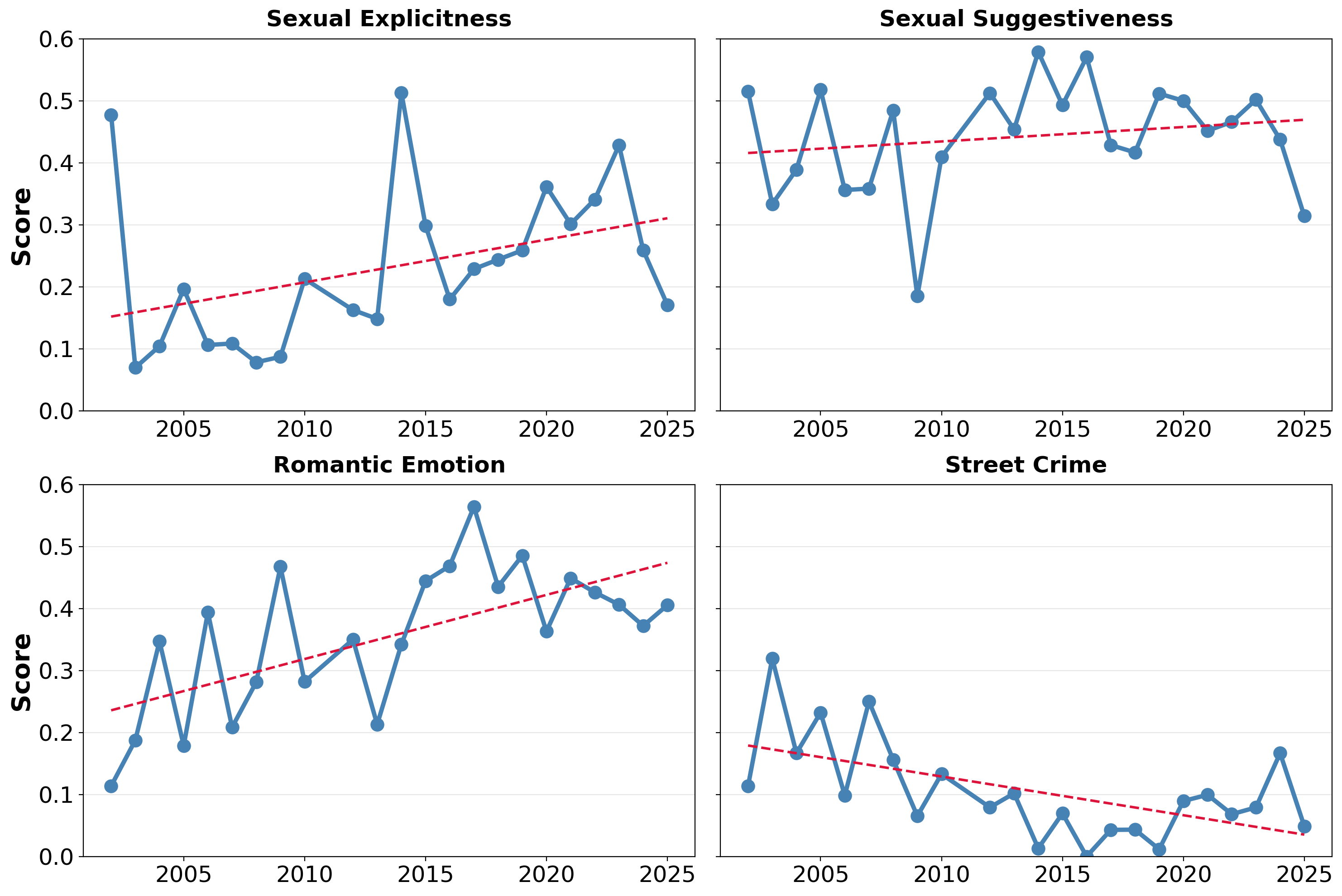}
\caption{Yearly mean (marker and line) and linear trend (dashed) for four
thematic dimensions, 2002--2025. Each marker averages all songs released in
that year, and the dashed line is a least-squares fit over those yearly means
rather than over individual songs. All panels share the same vertical scale.}
\label{fig:temporal}
\end{figure}

The corpus spans more than two decades, which lets us track how each dimension
evolves over time. A similar longitudinal approach was used by
\citet{arevalo2018diabla}, who manually coded 70 reggaeton songs (2004--2017)
across five types of gender violence: physical, sexual, economic, symbolic, and
psychological. Figure~\ref{fig:temporal} plots the yearly mean of four
dimensions, together with a linear trend fitted across years. These are the four
dimensions that carry the trends discussed below: sexual explicitness, sexual
suggestiveness, romantic emotion, and street crime. All four panels share a
common vertical scale, so the slopes can be compared directly.

Some of these trends run counter to the popular narrative around reggaeton.
\emph{Street crime} is often cited as emblematic of the genre's early
``gangsta'' aesthetic. Nevertheless, it declines over the period, with the
fitted value falling from 0.18 in 2002 to 0.04 in 2025 ($r=-0.57$, $p=0.005$).
\emph{Romantic emotion} shows the steepest increase of the four. It roughly
doubles, from 0.24 to 0.47 ($r=0.65$, $p=0.001$). This rise occurs alongside,
not instead of, the growth in explicit content discussed next. Among the
dimensions not plotted, \emph{substance use} also trends modestly upward
(0.10 to 0.24; $r=0.50$, $p=0.014$).

All the trends reported in this Section are corpus-level. They describe what
the corpus contains in each year, rather than how any individual artist
evolved. The distinction matters for two of them, as we discuss at the end of
the Section.

The two sexual dimensions diverge in a way that echoes the suggestive/explicit
distinction drawn in Step 2. \emph{Sexual explicitness} rises substantially,
roughly doubling from a fitted value of 0.15 to 0.31 ($r=0.39$, $p=0.065$).
\emph{Sexual suggestiveness}, instead, is comparatively flat (0.42 to 0.47;
$r=0.18$, $p=0.40$). The latter is not a statistically detectable trend at
conventional thresholds. In other words, the growth in sexual content documented
for this corpus is concentrated in \emph{explicitness}, rather than in a
broad-based increase in suggestiveness.

The composite sexual score (not plotted) reflects the same asymmetry. Comparing
songs from 2002--2012 against 2013--2025, the mean composite sexual score rises
from 0.288 ($n=328$) to 0.383 ($n=931$), a 33\% increase. This is a coarse
two-era split, chosen close to the corpus's temporal midpoint.

One caveat is worth remarking. The corpus does not contain the same artists in
every year. Therefore, part of
the trend may reflect which artists were active in later years rather than a
within-artist evolution. Nevertheless, the rise in \emph{sexual explicitness}
survives this concern. Restricting the variation to each artist's own catalogue
(an artist-demeaned regression), the score still increases by 0.064 per decade
(95\% CI $[0.028, 0.130]$, bootstrapped over artists). The same holds for
\emph{substance use} (0.049 per decade; 95\% CI $[0.008, 0.187]$). The trends in
\emph{romantic emotion} and \emph{street crime}, instead, do not survive it.
Once the variation is restricted to each artist's own catalogue, both slopes
are compatible with zero. These two trends therefore describe a change in who
was recording, rather than a change in what a given artist recorded.
Disentangling the two effects with a corpus designed for that purpose is a
natural target for follow-up work.

\subsection{Comparison with Spotify's explicit flag}
\label{sec:spotify-comparison}

As a final check, we compare the method's \emph{sexual explicitness} score
against Spotify's own \emph{explicit} flag for every song in the corpus.
Spotify states that it flags explicit content ``depending on information we
receive from
rights-holders''.\footnote{\url{https://support.spotify.com/us/article/explicit-content/},
accessed 8 August 2026.} The flag is therefore supplied by labels and
distributors, rather than produced by a content analysis of Spotify's own.
Moreover, no public criteria define what counts as explicit. The flag plausibly
covers profanity, drug references, or violence, in addition to (or instead of)
sexual content. This comparison should therefore be read as a check on coverage,
not as a ground-truth evaluation of the method.

\begin{figure}[H]
\centering
\includegraphics[width=\textwidth]{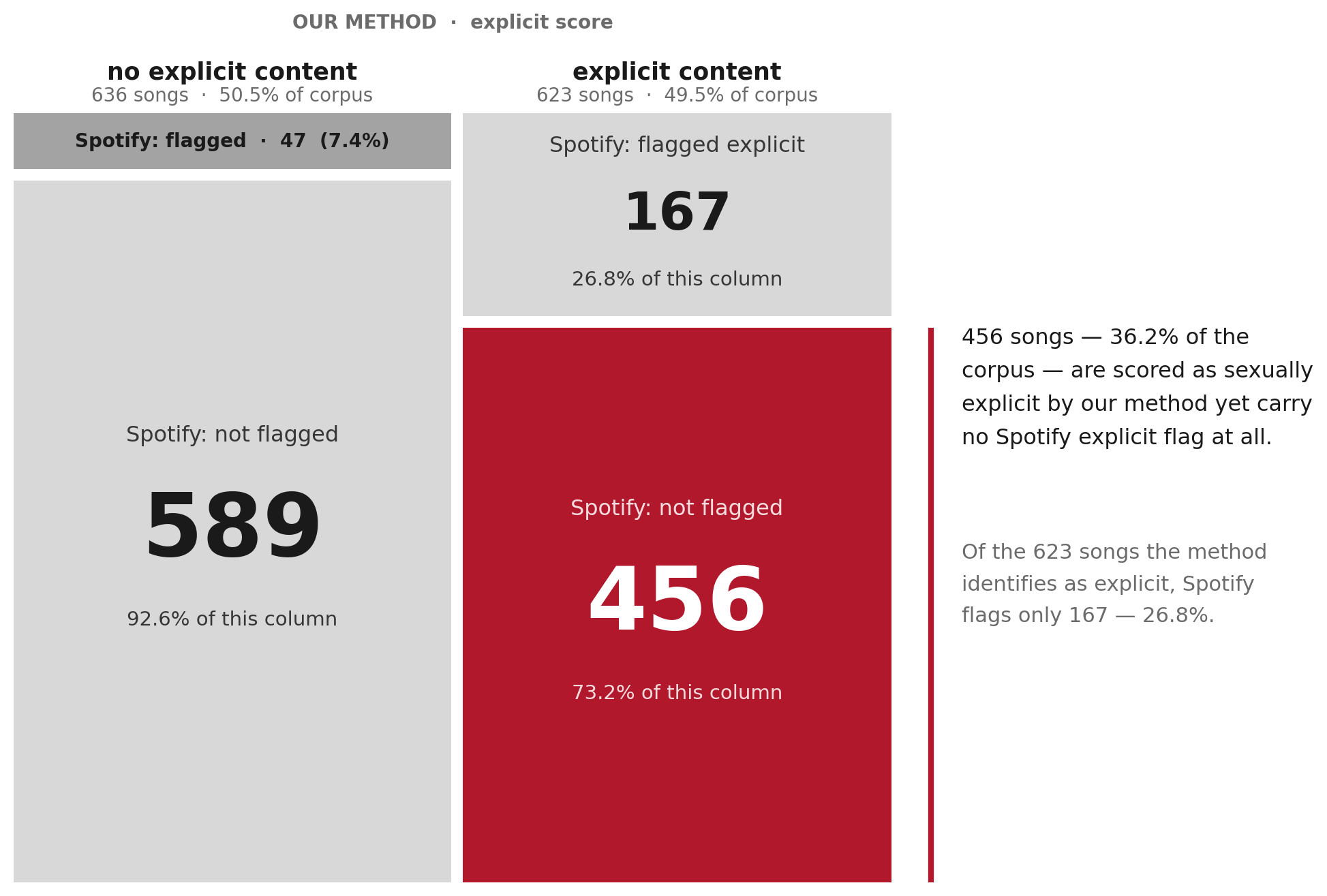}
\caption{Songs by agreement between Spotify's \emph{explicit} flag and the
method's sexual-explicitness score. Block area is proportional to the number of
songs. A song counts as explicit for the method when its sexual-explicitness
score is non-zero.}
\label{fig:spotify-confusion}
\end{figure}

Figure~\ref{fig:spotify-confusion} splits the corpus into four groups,
according to whether each source flags a song as explicit. We use the same
threshold as elsewhere in this paper ($\text{explicit}_{\text{norm}} > 0$), that
is, the model detected some explicit content at all.

The two quadrants where the sources agree account for the majority of the
corpus: 589 songs that neither source flags, and 167 songs that both flag. The
two disagreement quadrants, however, are far from symmetric.

Only 47 songs carry Spotify's explicit flag without a matching
sexual-explicitness score from our method. Spotify's flag is not restricted to
sexual content, so this quadrant most likely reflects songs flagged for
non-sexual reasons (profanity, drug references, violence), rather than a failure
of the method. The quadrant that matters is the
opposite one: 456 songs, or 36.2\% of the corpus, that the method scores as
sexually explicit but that carry no Spotify explicit flag at all. Put
differently, of the 623 songs our method identifies as containing explicit
sexual content, Spotify's flag catches only 167 (26.8\%). Nearly three out of
every four sexually explicit songs in this corpus go unflagged by Spotify.

We think this gap is the clearest practical illustration of why a method like
the one presented here is useful. A single binary flag with no published
criteria is a coarse instrument for something as specific as sexual
explicitness. Reggaeton is one of the most sexually explicit music genres by
reputation, and yet a large share of the explicit content in this corpus passes
through that flag unnoticed.

\citet{zamacola2026finetuning} arrive at a related conclusion from the detection
side. They propose an automated classifier together with a graded, age-based
rating scheme for streaming platforms, modeled on the PEGI system used for video
games. The measurements reported here are consistent with the premise behind
that proposal. Whatever Spotify's flag is tracking, it is not sexual
explicitness as our method scores it.

\section{Conclusion}
\label{sec:conclusion}

We presented a method for quantifying thematic content in song lyrics, and
applied it to reggaeton. The method takes a list of artists and returns a table
in which each row is a song and each column a thematic dimension scored by an
LLM. It proceeds in three steps: data collection with a manual review pass,
per-song scoring along nine dimensions, and corpus assembly. The scoring prompt
is calibrated against hand-labeled songs before being applied at scale, and we
treat that calibration as part of the method rather than a preliminary. We
applied the method to 1{,}259 songs by 12 artists released between 2002 and
2025.

The resulting corpus supports four analyses. In the first place, \emph{sexual
suggestiveness} is the most prevalent single dimension in the corpus, and it is
nearly twice as prevalent as \emph{sexual explicitness}. This asymmetry supports
keeping the two apart, rather than collapsing them into one measure. Secondly,
artists differ widely. The mean composite sexual score ranges from 0.14 to 0.56
across the 12 artists, and their profiles across the nine dimensions read as
distinct archetypes rather than as variations in intensity along a single axis.

Thirdly, over the two decades covered, the rise in sexual content is
concentrated in explicitness rather than in suggestiveness. It occurs alongside
a simultaneous rise in romantic emotion and a decline in street crime. Finally,
of the songs the method scores as containing explicit sexual content, only
roughly one in four carries Spotify's \emph{explicit} flag.

None of this is specific to sexual content. The dimensions live in the scoring
prompt, so the same procedure can be pointed at a different coding scheme.
Candidates are the six themes of \citet{bretthauer2007feminist}, the five types
of gender violence coded by \citet{arevalo2018diabla}, or the hostile and
benevolent sexism of \citet{glick1996ambivalent}. A researcher can equally
define their own categories for their own question. It can also be pointed at a different genre or
language, each with its own calibration cycle. Section~\ref{sec:method}
discusses the limitations that come with this, and the directions we see for
making the procedure more accurate and more automatic.

The findings reported here are exploratory, and they describe one corpus. What
we offer for reuse is the procedure. The data collection code, the scoring
prompt, and the corpus are released, so that these numbers can be reproduced,
audited, or recomputed under a different scoring criterion. The bibliometric
review cited in Section~\ref{sec:introduction} notes that the instruments used
to study reggaeton remain largely unconsolidated. An explicit, documented,
reusable scoring procedure is one way to address that, and we hope this is a
useful step in that direction.

\section*{Acknowledgments}

I thank my wife Ana for her valuable advice and her criteria, which improved
the research and the clarity of the visualizations.

\bibliographystyle{plainnat}
\bibliography{references}

\end{document}